\documentclass[reprint,amsmath,amssymb,aps,noshowpacs,superscriptaddress]{revtex4-2}

\usepackage{graphicx}
\usepackage{dcolumn}
\usepackage{bm}
\usepackage{mathtools}
\usepackage[left=15mm,right=15mm,top=30mm,columnsep=15pt]{geometry} 
\usepackage{xcolor}
\usepackage{bbold}
\usepackage{svg}
\usepackage{siunitx}
\usepackage{braket}
\usepackage[T1]{fontenc}
\usepackage[utf8]{inputenc}
\usepackage[english]{babel}
\usepackage{hyperref}
\hypersetup{colorlinks=true, citecolor=blue, urlcolor=blue, linkcolor=blue}
\usepackage{booktabs}
\newcommand{\ra}[1]{\renewcommand{\arraystretch}{#1}}

\begin{document}

\title{Feedback cooling a levitated nanoparticle's libration to below 100 phonons}

\author{Jialiang~Gao}
\thanks{These authors contributed equally.}
\affiliation{Photonics Laboratory, ETH Zürich, 8093 Zürich, Switzerland}
\affiliation{Quantum Center, ETH Zürich, Zürich, Switzerland}

\author{Fons~van der Laan}
\thanks{These authors contributed equally.}
\affiliation{Photonics Laboratory, ETH Zürich, 8093 Zürich, Switzerland}
\affiliation{Quantum Center, ETH Zürich, Zürich, Switzerland}
\affiliation{Center for Nanophotonics, AMOLF, 1098 XG Amsterdam, The Netherlands}

\author{Joanna A.~Zielińska}
\affiliation{Photonics Laboratory, ETH Zürich, 8093 Zürich, Switzerland}
\affiliation{Quantum Center, ETH Zürich, Zürich, Switzerland}

\author{Andrei~Militaru}
\affiliation{Photonics Laboratory, ETH Zürich, 8093 Zürich, Switzerland}
\affiliation{Quantum Center, ETH Zürich, Zürich, Switzerland}

\author{Lukas~Novotny}
\affiliation{Photonics Laboratory, ETH Zürich, 8093 Zürich, Switzerland}
\affiliation{Quantum Center, ETH Zürich, Zürich, Switzerland}

\author{Martin~Frimmer}
\affiliation{Photonics Laboratory, ETH Zürich, 8093 Zürich, Switzerland}
\affiliation{Quantum Center, ETH Zürich, Zürich, Switzerland}

\date{\today}

\begin{abstract}
Macroscopic rotors are interesting model systems to test quantum theory and for quantum sensing. A promising approach for bringing these systems to the quantum regime is to combine sensitive detection with feedback cooling to reduce the thermal occupation of the mechanics. Here, we implement a backward-scattering scheme to efficiently detect all three libration modes of an optically levitated nanoparticle. We demonstrate parametric feedback cooling of all  three libration degrees of freedom to below 16~mK, with one of the modes reaching the temperature of 1.3~mK, corresponding to a mean phonon number of 84. Finally, we characterize the backward-scattering scheme by determining its measurement efficiency to be 0.5\%.
\end{abstract}
\maketitle

\section{Introduction} \label{sec:Introduction}

Levitodynamics is the field of controlling levitated macroscopic objects. In absence of a mechanical clamping mechanism, their motion can be highly isolated from the environment. Consequently, this platform is a prime candidate for studying macroscopic quantum dynamics as well as for sensing small forces and torques~\cite{gonzalez2021levitodynamics}. A milestone towards demonstrating quantum phenomena in levitation was reached by cooling the center-of-mass (c.m.) motion of an optically trapped particle to its ground state, both via coherent scattering into an optical cavity~\cite{delic2020cooling, ranfagni2022two, piotrowski2023simultaneous} and via measurement-based feedback~\cite{tebbenjohanns2021quantum, magrini2021real, kamba2023revealing}. 

Besides studying translational motion, levitated anisotropic particles are especially suited for exploring the control of rotational degrees of freedom. 
Rotational motion is particularly enticing in the quantum regime since it is intrinsically nonlinear, and allows for unique quantum interference effects such as orientational quantum revivals~\cite{stickler2018probing}. Remarkable experimental progress has been made in rotational levitodynamics. On the one hand, levitated nanoparticles have been driven into rotation at GHz frequencies~\cite{reimann2018ghz, ahn2018optically, ahn2020ultrasensitive, ju2023near}, and fluctuations of rotation rate have been controlled~\cite{van2020optically, blakemore2022librational,kuhn2017optically}. 
On the other hand, a restoring torque on the orientation creates libration modes, which behave like harmonic oscillators for small angular displacements. 
Cooling several libration modes simultaneously has been successfully demonstrated using a cavity~\cite{pontin2023simultaneous} as well as measurement-based feedback~\cite{bang2020five, kamba2023nanoscale}. 
Furthermore, measurement backaction has been observed on optically levitated rotors~\cite{van2021sub}. 
However, a detection scheme with a sufficiently high efficiency for quantum control is still to be demonstrated~\cite{tebbenjohanns2019optimal}. 

While the first wave of experiments in rotational levitodynamics used cylindrically symmetric rotors (termed nanodumbbells), it became clear that they suffer from degenerate libration modes and their uncontrolled coupling by thermally driven spinning~\cite{seberson2019parametric,bang2020five, van2021sub, zielinska2023spinning}.
As a result, the attention of the community has recently shifted towards controlling all six degrees of freedom of fully anisotropic particles with three distinct moments of inertia~\cite{pontin2023simultaneous, kamba2023nanoscale}.
These rotors exhibit nondegenerate libration modes, making each mode individually addressable in the frequency domain. 
A particularly tantalizing example of a rotational quantum effect (without counterpart in translational motion) requiring such an anisotropic rotor are quantum tennis-racket flips~\cite{ma2020quantum}. 
One requirement for their observation is sufficient cooling of the libration modes. 
Interestingly, recent experiments underline that control of librational motion is also key to sensitive experiments on the c.m.\ motion of levitated particles~\cite{kamba2023revealing}.
The further development of detection and cooling schemes for librational motion, especially of fully anisotropic levitated rotors, is thus of significant relevance to the entire levitodynamics community. 

\begin{figure*}
    \centering
    \includegraphics[width = 0.8\textwidth]{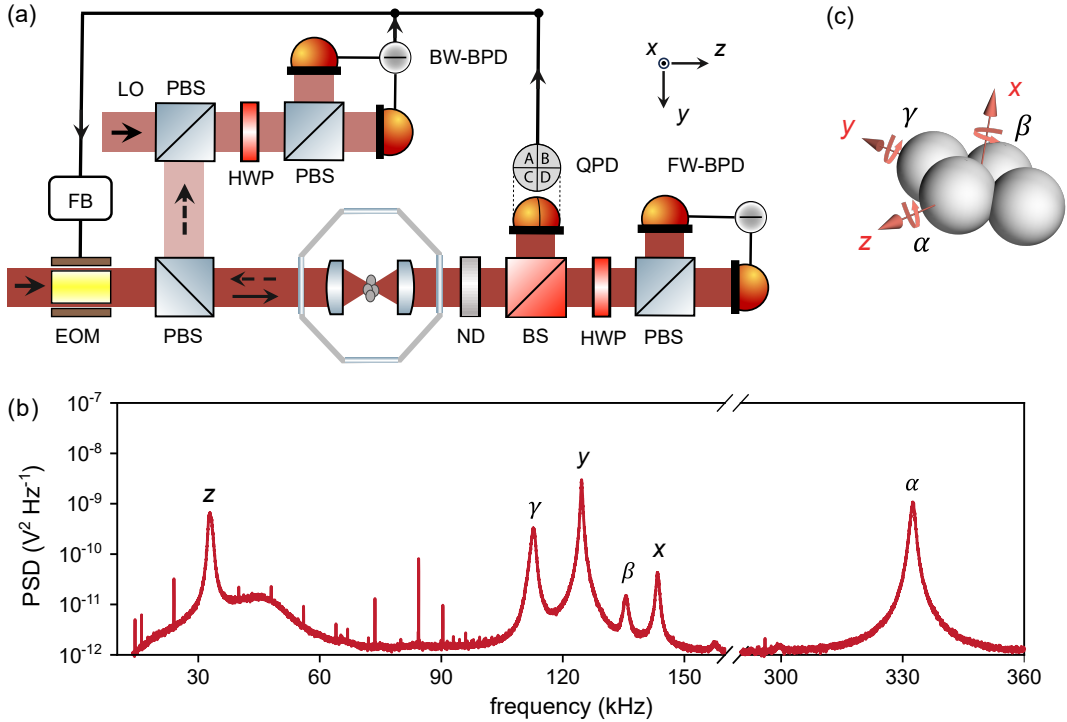}
    \caption{{(a)} Schematic of the experimental setup. Inside a vacuum chamber, we generate an optical trap by strongly focusing a laser beam (propagating along $z$ and linearly polarized along $y$). In the forward direction, we detect the particle's c.m.\ motion on a quadrant photodetector (QPD) and libration motion on a balanced photodetector (FW-BPD). In the backward direction, the scattered light from the particle is overlapped with a local oscillator (LO) on another balanced photodetector (BW-BPD) for efficient homodyne detection of the libration modes. We use the signals from QPD and BW-BPD to modulate the power of the laser with an electro-optic modulator (EOM) for feedback cooling c.m.\ and librational motion, respectively. 
    {(b)} Power spectral density (PSD) of the signal measured by FW-BPD at $0.8\,$mbar. The six labeled peaks correspond to the particle's c.m.\ ($x,y,z$) and libration ($\alpha,\beta,\gamma$) modes, respectively. The additional spikes are electronic noise from our detectors and data acquisition card. 
    {(c)} The orientation of the particle is confined with its long axis along the polarization ($y$) axis and its mid-axis along the laser propagation ($z$) axis. Three libration modes $\alpha$, $\beta$ and $\gamma$ rotate the particle around the $z$, $x$, and $y$ axes, respectively. 
    }
    \label{fig:setup}
\end{figure*}

In this work, we use measurement-based parametric feedback cooling to control the orientation of a nanoparticle with three distinct moments of inertia.  
All three non-degenerate libration modes are cooled to below $16$~mK, with the temperature of the coldest mode reaching $1.3$~mK, corresponding to a mean phonon occupation of 84. 
Our approach relies on a back-scattering measurement scheme to enhance the detection efficiency by three orders of magnitude compared to its forward scattering counterpart. 
Additionally, we experimentally determine the measurement efficiency $\eta=0.5\%$ for the coldest libration mode in our system.

\begin{figure*}
    \centering
    \includegraphics[width = 0.85\textwidth]{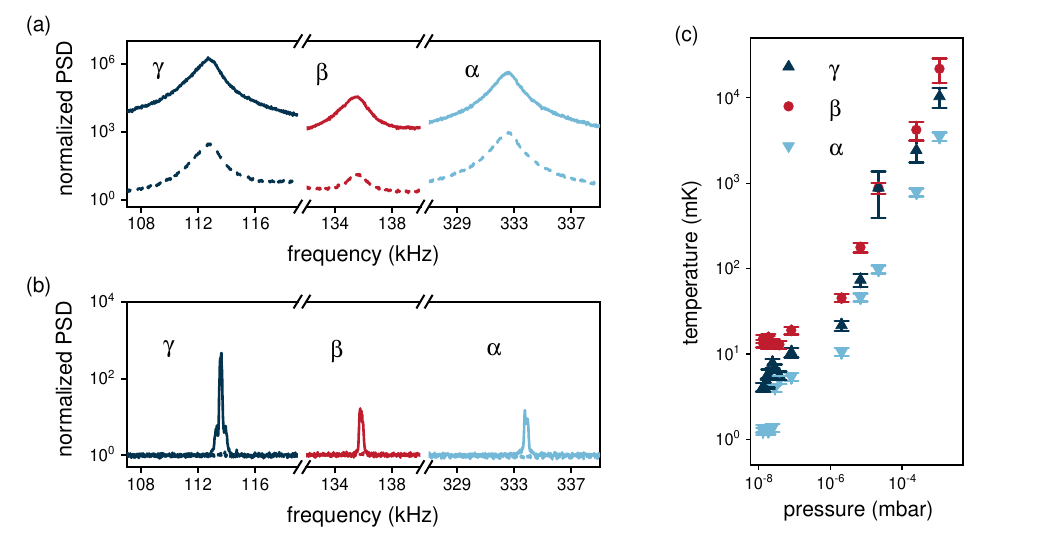}
    \caption{{(a)} Comparison of forward and backward detection. Normalized power spectral densities (PSD) of the libration modes $\alpha$, $\beta$, and $\gamma$ at $0.8\,$mbar. The spectra recorded on the backward detector (BW-BPD, solid lines) and forward detector (FW-BPD, dashed lines) are normalized to their respective noise floor in order to compare their signal-to-noise ratios. 
    {(b)} Normalized power spectral densities of the same libration modes as in (a) under feedback cooling at $3\times10^{-8}\,$mbar. 
    {(c)} Temperature of the three libration modes as a function of pressure measured under parametric feedback cooling. At $3\times10^{-8}$~mbar, the settings of the feedback for the $\alpha$ mode were optimized.}
    \label{fig:lib_coolingl}
\end{figure*}

\section{Experimental setup}\label{sec:Setup}

We show our experimental setup in Fig.\,\ref{fig:setup}(a). The trapping beam
[wavelength $1550\,$nm, optical power ${900(50)\,\mathrm{mW}}$], propagates along the $z$ axis and is linearly polarized along the $y$ axis. Inside a vacuum chamber, we form an optical trap by strongly focusing the beam with an aspheric lens (NA=0.69). 
We trap a particle in the focus and collect its scattered light in the backward and forward direction using the trapping lens and an identical lens, respectively.  
In the forward direction, the scattered light, together with the trapping beam, is recollimated and half of the light is sent to a quadrant photodetector (QPD) for c.m.\ motion detection. For libration detection, the other half is distributed on a balanced photodetector (FW-BPD) using a half-wave plate (HWP) and polarizing beam splitter (PBS). In this forward detection scheme, the trapping beam is automatically overlapped with the scattered light and acts as a local oscillator. This makes the detection scheme robust against drift and straightforward to implement.
However, we have to attenuate the light (and therefore also the signal) on the detectors with a neutral density filter (ND, transmission $0.2$\%) to avoid damage to the detectors from the high power of the trapping beam.
A spectrum recorded by the FW-BPD at 0.8~mbar is shown in Fig.~\ref{fig:setup}(b). The spectrum shows three c.m.\ modes ($x$, $y$, $z$) and three libration modes ($\alpha$, $\beta$, $\gamma$), indicating a trapped particle with three distinct moments of inertia.
To increase the detection efficiency of the libration mode in the polarization plane of the optical trap (characterized by the angle $\alpha$), we implement an additional back-scattering detection scheme as proposed by Ref.~\cite{tebbenjohanns2022optimal}.
To this end, in the backward direction, we select the $x$-polarized scattered light using a PBS and overlap the light with a local oscillator 
(LO, optical power $\sim 5\,$mW) in a heterodyne measurement using another HWP, PBS, and balanced photodetector (BW-BPD). The frequency of the LO differs by only 9~Hz from that of the trapping beam. Since this frequency difference is much smaller than the linewidths of the motional peaks, the motional sidebands are effectively overlapped in a single-sided power spectral density without significant additional broadening. With this method, we circumvent having to phase lock the LO at the cost of halving the signal strength. The back-scattering scheme requires no light attenuation, since we independently control the power of the LO.

To stabilize the particle inside the trap, we perform parametric feedback cooling on all translational and librational degrees of freedom using the signals from the QPD (for c.m.\ motion) and BW-QPD (for libration)~\cite{seberson2019parametric,gieseler2012subkelvin}. For each mode, a phase-locked loop (PLL) tracks frequency and phase from which we generate a feedback signal at twice the oscillation frequency. We apply the sum of all feedback signals to an electro-optic modulator (EOM), resulting in a modulation of the trapping beam power which cools all modes simultaneously~\cite{jain2016direct, vovrosh2017parametric, van2021sub}.

We load silica spheres (nominal diameter $156\,$nm, dispersed in isopropanol) into the trap using a nebulizer. The anisotropic particles with three distinct moments of inertia are assembled from the spheres. Even though we cannot control the exact shape of each trapped particle, we are able to repeatably trap particles with three distinct moments of inertia by controlling the concentration of silica spheres in the initial solution. The exact motional frequencies slightly differ from particle to particle. All experiments in this work are performed with the same particle. 

Let us turn our attention to the orientation of the particle in the trap. The observation of three libration modes in Fig.~\ref{fig:setup}(b) indicates that the orientation is fully confined in three dimensions. 
Note that, when trapped in a purely linearly polarized light field, the orientation of an anisotropic point scatterer is expected to exhibit only two confined libration modes (along with one unconfined degree of freedom corresponding to free rotation around the long axis)~\cite{bang2020five, seberson2019parametric, van2022rotational, zielinska2023controlling}. We conjecture that the complete 6D confinement observed in this work is a consequence of the additional orientation confinement mechanism due to the intensity gradient of the trap~\cite{zeng2023gradient, kamba2023nanoscale}. Following the arguments presented in~\cite{kamba2023nanoscale}, we expect the combination of the trapping beam's polarization and intensity gradient to result in the longest axis of the particle aligning along the polarization ($y$) axis, and the mid-axis along the laser propagation ($z$) axis, as illustrated in Fig.~\ref{fig:setup}(c). We verify the particle orientation by measuring the c.m.\ damping rates $\gamma_i$ ($i\in\{x,y,z\}$). For their ratios, we find $(\gamma_x:\gamma_z:\gamma_y = 1.7:1.3:1)$. The order of the damping ratios $\gamma_x>\gamma_z>\gamma_y$ indicates that, indeed, the particle orientation is as depicted in Fig.~\ref{fig:setup}(c).

\section{Libration detection and cooling} \label{sec:Lib cooling}

Having identified the three libration modes, we turn our attention to comparing the forward and backward detection schemes. Since the optical powers on the FW-BPD and the BW-BPD are unequal and the detectors have different transimpedance gains, we normalize the obtained libration spectra to their respective noise floors. In Fig.~\ref{fig:lib_coolingl}(a), we show the power spectral densities of the $\alpha$ (light blue), $\beta$ (red), and $\gamma$ (dark blue) modes recorded on the BW-BPD at $0.8$\,mbar. As dashed lines, we show the corresponding modes simultaneously recorded on the FW-BPD. 
We observe that the signal-to-noise ratio of the BW-BPD exceeds that of the FW-BPD by approximately three orders of magnitude for all libration modes. We mainly attribute this increase in signal to the reduction of light attenuation in the backward direction in comparison to the forward direction, where the ND filter introduces a loss of signal of three orders of magnitude. 

We continue by exploiting the signal detected by the BW-BPD to perform PLL feedback cooling on all three libration modes between $10^{-3}$ and $10^{-8}$~mbar. Figure~\ref{fig:lib_coolingl}(b) shows the power spectral densities for the $\alpha$, $\beta$, and $\gamma$ modes at $3\times10^{-8}$\,mbar. Again, the solid lines are recorded using the BW-BPD and the dashed lines using the FW-BPD.  By integrating the area under the peak of each mode and subtracting the noise floor, we extract the effective temperature of each libration mode. 
Figure~\ref{fig:lib_coolingl}(c) shows the measured temperature as a function of pressure of the three libration modes under feedback cooling. For pressures above $10^{-7}$~mbar, the temperatures of all three modes decrease linearly with pressure. However, between $10^{-8}$ and $10^{-7}$~mbar the temperatures of the $\alpha$ and $\beta$ modes level off, indicating that they are no longer limited by pressure~\cite{jain2016direct}. Note that, at $3\times10^{-8}$~mbar, the settings of the feedback for the $\alpha$ mode were optimized. 
We summarize the measured temperatures and mean phonon occupation numbers at the lowest pressure of $1.3\times10^{-8}$\,mbar in Table.~\ref{tab:temperatures}. The backward detection scheme is optimized for the $\alpha$-mode (libration in the focal plane of the optical trap) and we indeed find the lowest phonon occupation of 84 for that mode. We stress that, under feedback cooling at low pressures, the libration peaks  are not discernible above the noise floor on the FW-BPD, as seen in Fig.~\ref{fig:lib_coolingl}(b). Thus, feedback cooling using the forward detection would not have been able to reach similar temperatures.

\begin{table}\centering
\ra{1.2}
\begin{tabular}{p{0.15\linewidth} p{0.4\linewidth} p{0.25\linewidth}}
    \toprule[0.7pt]
     \textbf{Mode} & \textbf{Temperature (mK)} & \textbf{Occupation}\\
    \midrule[0.3pt]
    $\alpha$ & $1.34\pm 0.14$ & $84\pm9$\\
    $\beta$ & $15\pm2$ & $2298\pm248$\\
    $\gamma$ & $4.1\pm0.5$ & $742\pm87$\\
    \bottomrule[0.7pt]
\end{tabular}
\caption{Measured temperatures and mean phonon occupations for each libration mode at $1.3\times10^{-8}$\,mbar.}
\label{tab:temperatures}
\end{table}

\section{Measurement efficiency}\label{sec:Reheating}

Let us discuss the cooling limits of our backward detection scheme. 
The limitation to cooling performance in our experiment primarily stems from the parametric feedback method itself, rather than the detection efficiency~\cite{jain2016direct}. The phonon occupation can be further lowered by applying linear feedback based on the same backward detection scheme.
The performance of linear feedback cooling is limited by the measurement efficiency $\eta \leq 1$, which can be defined via the imprecision-backaction product $\tilde{S}_{\tau\tau} \tilde{S}_\mathrm{imp} \eta = \hbar^2$, where $\tilde{S}_{\tau\tau}$ is the single-sided power spectral density with real frequencies of the fluctuating torque driving the libration mode, $\tilde{S}_\mathrm{imp}$ is the measurement imprecision of the orientation angle of the librator, and $\hbar$ is the reduced Planck constant~\footnote{We define our power spectral densities according to $\langle \alpha^2 \rangle = \int_0^\infty \tilde{S}_{\alpha\alpha}(f)\,\mathrm{d}f$}. Thus, the measurement efficiency quantifies how far a measurement is from the Heisenberg limit of the imprecision-backaction product. Both $\tilde{S}_{\tau\tau}$ and $\tilde{S}_\mathrm{imp}$ can be directly measured, such that we can experimentally determine the measurement efficiency. The measurement imprecision is directly given by $\tilde{S}_\mathrm{imp} = \tilde{S}_\mathrm{imp}^\mathrm{exp} / c^2$, where $\tilde{S}_\mathrm{imp}^\mathrm{exp}$ is the detector noise floor in $\mathrm{V^2/Hz}$ and $c$ is the calibration factor in $\mathrm{V/rad}$, which is determined from the equipartition theorem ${c^2 = I \Omega_\alpha^2 \langle v_\mathrm{cal}^2 \rangle/(k_\mathrm{B} T_\mathrm{cal})}$, with $k_B$ the Boltzmann constant. Here, $\langle v_\mathrm{cal}^2 \rangle$ is found during a calibration measurement at known temperature $T_\mathrm{cal}$ by integrating the measured power spectral density (units $\mathrm{V^2/Hz})$~\cite{Hebestreit}. In order to determine $\tilde{S}_{\tau\tau}$, we perform reheating measurements of libration mode $\alpha$ at different pressures~\cite{van2021sub}. At the start of each iteration of the reheating protocol, feedback cooling of the mode $\alpha$ is turned off for $0.6$~s, after which feedback cooling is turned on again. We demodulate the libration signal with a bandwidth of $4$~kHz with a lock-in amplifier to continuously measure the energy of the libration mode (in units of $\mathrm{V^2}$). 
\begin{figure}
    \centering
    \includegraphics[width = 0.5\textwidth]{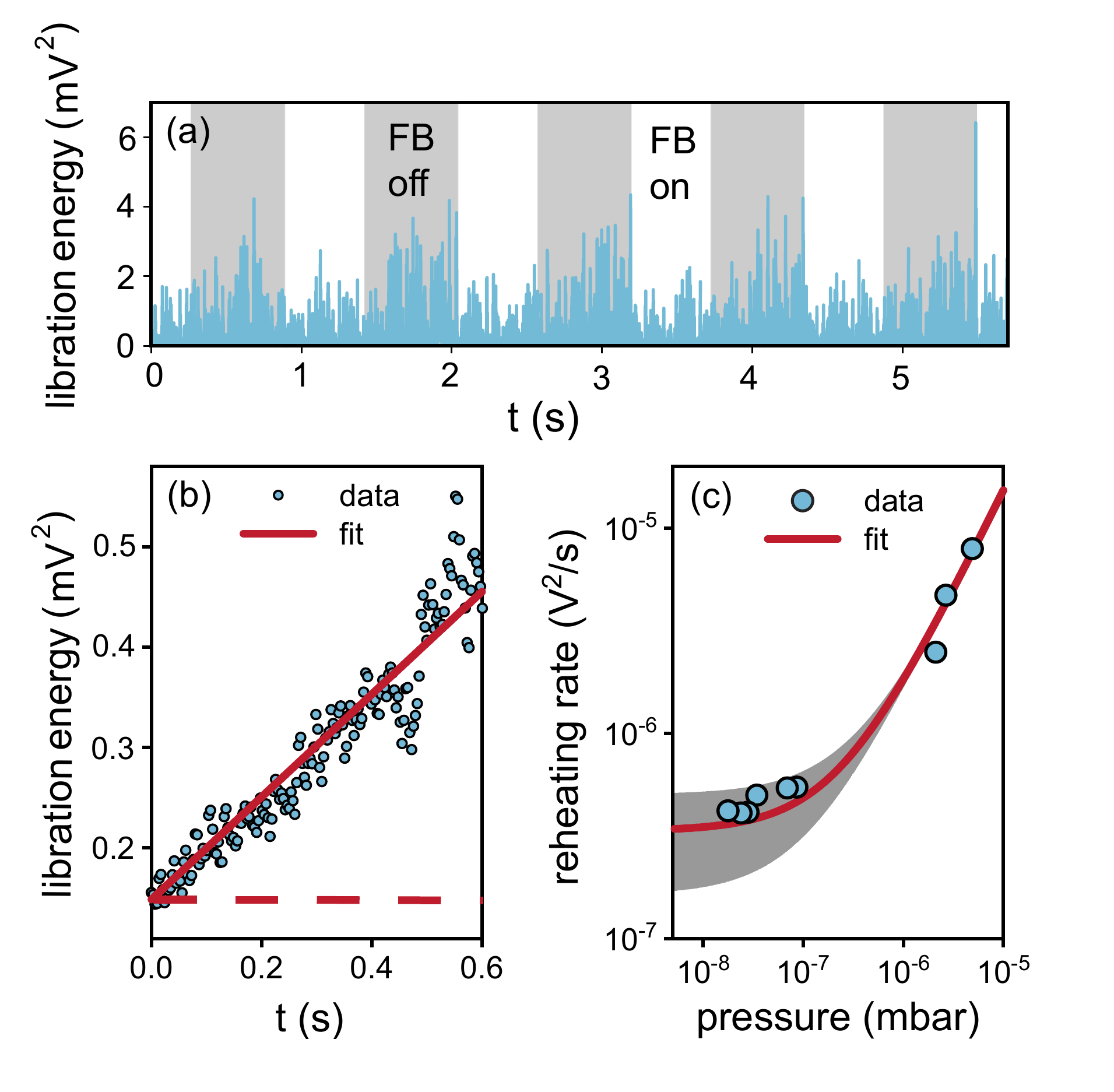}
    \caption{{(a)} During the reheating protocol the feedback cooling is toggled on (white areas) and off (gray areas), while we measure the energy  of the $\alpha$ mode in mV$^2$. 
    {(b)}~Energy (in units of detector signal variance mV$^2$) as a function of time at $7\times10^{-8}$\,mbar after switching off the parametric feedback cooling. The data is averaged over all 50 cycles. A linear fit to the data is shown as the red solid line. The average  energy under continuous feedback cooling is shown as the dashed line. 
    {(c)}~Heating rate (blue circles) as a function of pressure. A linear fit to the data is shown as the red solid line. The shaded area represents one standard deviation of the fit.}
    \label{fig:Reheating}
\end{figure}
Part of a typical timetrace of the libration energy is shown in Fig.~\ref{fig:Reheating}(a). 
The protocol is repeated 50 times and the stochastic reheating trajectories are averaged, as shown in~Fig.~\ref{fig:Reheating}(b). As expected, for timescales much shorter than the damping time, the energy increases linearly in time. We extract the heating rate $\Gamma^\mathrm{exp}$ (in units of $\mathrm{V^2/s}$) from a linear fit shown as the solid line in Fig.~\ref{fig:Reheating}(b). We constrain the fit at $t=0$ to be equal to the average energy under continuous feedback cooling, indicated by the dashed line. In Fig.~\ref{fig:Reheating}(c) we plot the heating rate measured at different pressures. As expected, the heating rate decreases linearly with pressure until it saturates at $10^{-8}$~mbar~\cite{van2021sub}. We fit $\Gamma^\mathrm{exp} = a \times p_\mathrm{gas} + \Gamma_\mathrm{res}^\mathrm{exp}$ to our data with the scaling $a$ and residual damping rate $\Gamma_\mathrm{res}^\mathrm{exp}$ as fit parameters. The fit and corresponding uncertainty are depicted as a solid line and shaded region, respectively. Using the fit, we determine the heating rate $\Gamma^\mathrm{exp}=3.5\times10^{-7}~\mathrm{V^2/s}$ at $1.3\times10^{-8}$~mbar. As a last step, we apply the fluctuation-dissipation theorem to find $\tilde{S}_\mathrm{\tau\tau} = 4 I^2 \Omega_\alpha^2 \Gamma^\mathrm{exp} / c^2$, where $I$ is the moment of inertia corresponding to mode $\alpha$, and $\Omega_\alpha$ its eigenfrequency. By rewriting the imprecision-backaction product, we find the measurement efficiency expressed in experimentally measurable quantities
\begin{align}
\label{eq:eta}
    \eta &= \left( \frac{\hbar}{2} \right)^2\frac{\Omega_\alpha^2 \langle v_\mathrm{cal}^2 \rangle^2}{(k_\mathrm{B} T_\mathrm{cal})^2} \frac{1}{\Gamma^\mathrm{exp} \tilde{S}_\mathrm{imp}^\mathrm{exp}}.
\end{align}
Note that the 
moment of inertia does not appear in Eq.~\eqref{eq:eta}. 
Thus, we can determine the measurement efficiency without the knowledge of the exact shape of the trapped particle.
For our current setup, we extract the measurement efficiency for the $\alpha$ mode to be $\eta = 0.5\%$. The maximum achievable detection efficiency using a lens with $\mathrm{NA}=0.69$ is $\eta_\mathrm{det} = 20\%$~\cite{tebbenjohanns2022optimal}. We mainly attribute the difference to optical loss of the backscattered light due to poor mode overlap with the Gaussian mode of the fiber-coupled detector, sweeping the phase of the LO instead of locking it, and contributions of electronic noise.

\section{Conclusion} \label{sec:Conclusion}

In conclusion, we have parametrically feedback cooled all three libration modes of an optically levitated anisotropic nanoparticle from room temperature to below $16$~mK. In particular, we obtained a record-low phonon occupation of 84 phonons for the libration mode in the focal plane of the optical trap. In order to reach this cooling performance, we implemented a backward scattering detection scheme. Even though the scheme is optimized for detecting libration in the focal plane, the other two libration modes are also clearly visible. We have observed an increase in detection efficiency of approximately three orders of magnitude for all libration modes for the backward detection scheme, compared with the commonly used forward scattering scheme. 

Moreover, using reheating measurements, we have extracted the measurement efficiency of our system to be $\eta=0.5$\%. 
Regarding the theoretical cooling limit, there is currently no theoretical framework linking the final temperature reachable using PLL feedback cooling to the measurement efficiency $\eta$. 
Nonetheless, for linear feedback cooling, which has been demonstrated for librational motion~\cite{bang2020five,kamba2023nanoscale}, one can cool a harmonic oscillator to a mean phonon occupation $\bar{n}_\mathrm{min}=(1/\sqrt{\eta}-1)/2$. Inserting the measurement efficiency of our system, we find $\bar{n}_\mathrm{min}=6.4$. 
We thus expect that, once amended with linear feedback control, our system will be able to perform sideband thermometry of a libration mode~\cite{purdy2015optomechanical, delic2019cavity,tebbenjohanns2020motional}. Moreover, by carefully minimizing the optical losses, ground state cooling of the libration mode may be within reach.

The backward detection scheme implemented in this work, together with the anisotropic nature of the trapped nanoparticle, allowed for feedback cooling to a level that represents a milestone towards rotational quantum experiments~\cite{stickler2021quantum}. For example, one requirement to observe quantum tennis-racket flips is to prepare the initial state of a nanoparticle according to the relation $\hbar\omega/k_BT\geq0.1$, where $\omega$ is the angular velocity of rotation around the mid-axis of the nanoparticle, and $T$ is the effective temperature of the libration motion along the mid-axis~\cite{ma2020quantum}.  For the nanoparticle cooled as described in this study, achieving the required initialization involves a moderate mid-axis rotation rate in the MHz range, which is within experimental reach~\cite{reimann2018ghz, ahn2018optically}. 

\vspace{5mm}

We acknowledge our colleagues at the Photonics Laboratory at ETH Zürich for useful discussions. This research was supported by the ETH Research Grant (Grant No. ETH-47 20-2) and European Union’s Horizon 2020 research and innovation programme under grant nos. 863132 (iQLev).

\bibliographystyle{apsrev4-2}
\bibliography{main}

\end{document}